\documentclass[particles,article,accept,moreauthors,pdftex,10pt,a4paper]{Definitions/mdpi} 

       \def\kB {k\llap{/\kern1pt}}
        \def\pB {p\llap{/\kern1pt}}
        \def\qB {q\llap{/\kern1pt}}
        \def\KB {K\llap{/\kern1pt}}
        \def\PB {P\llap{/\kern1pt}}
        \def\QB {Q\llap{/\kern1pt}}
        \def\FB {F\llap{/\kern1pt}}
        \def\SigmaB {\Sigma\llap{/\kern1pt}}

\usepackage{color,soul}

\usepackage{booktabs} 
\usepackage{multirow}
\usepackage{soul} 
\usepackage{microtype}
\setitemize{parsep=6pt,itemsep=0pt,leftmargin=*,labelsep=5.5mm}
\setenumerate{parsep=6pt,itemsep=0pt,leftmargin=*,labelsep=5.5mm}

\firstpage{676} 
\pubvolume{3}
\issuenum{4}
\articlenumber{44}
\pubyear{2020}
\copyrightyear{2020}
 \history{Received: 15 July 2020; Accepted: 7 October 2020; Published: 20 October 2020}
\updates{yes} 

\continuouspages{yes}

\Title{Direct Photons from Hot Quark Matter in Renormalized Finite-Time-Path QED}

\Author{Ivan Dadi\'c $^{1,*}$,  Dubravko Klabu\v{c}ar $^{2}$\orcidA{} and  Domagoj Kui\'c $^{1}$}

\AuthorNames{Ivan Dadi\'c ,  Dubravko Klabu\v{c}ar and  Domagoj Kui\'c}

\address{%
$^{1}$ \quad Rudjer Bo\v skovi\'c Institute, P.O. Box 180, 10002 Zagreb, Croatia; Domagoj.Kuic@irb.hr\\
$^{2}$ \quad Physics Department, Faculty of Science-PMF, University of Zagreb, Bijeni\v{c}ka c. 32, 
10000 Zagreb, Croatia; klabucar@phy.hr}

\corres{Correspondence: dadic@irb.hr}

\abstract{Within the finite-time-path out-of-equilibrium quantum field theory (QFT), we calculate direct photon emission from early stages of heavy ion collisions, from
 a narrow window, in which uncertainty relations are still important and they provide a new
 mechanism for production of photons. The basic difference with respect to earlier
 calculations, leading to diverging results, is that we use renormalized QED of quarks and photons. Our result is a finite contribution that is consistent with uncertainty relations.
}


\begin{document}




\section{Introduction}

Heavy ion collisions (HIC) result in many-particle final states that carry a lot of information, which is not easy to decode~\cite{Baym:2016wox,Pasechnik:2016wkt,David:2019wpt}, including, for example, the~recently emerged ‘direct photon puzzle’~\cite{David:2019wpt}.
Various attempts to describe HIC by theoretical means include increasingly involved
approaches, such as S-matrix QFT and equilibrium, as well as out-of-equilibrium QFT~\cite{Schwinger:1960qe,Keldysh:1964ud,KadanoffBook,Danielewicz:1982kk,Chou:1984es,Rammer:1986zz,Landsman:1986uw,Calzetta:1986cq,Niemi:1987pm,Remler:1990,LeBellacBook,Brown:1998zx,Blaizot:2001nr}.
Finally, the~fast evolving nature of HIC requires the finite-time-path description.
Pertinent calculations~\mbox{\cite{bv,bvhs,devega,bvs0,wbvl,wb,ng,bv68,bvhep}} of
production of photons from the early stage of HIC, and~the heated discussion
 afterwards~\cite{Arleo:2004gn}, with~the criticism on infinite energy that is
 released in photon yield, indicate that the subject is far from~settled.
 

This calculation is performed in the finite-time-path (FTP) out-of-equilibrium QFT.
FTP is a variation of out-of-equilibrium QFT with propagators defined through a finite-time contour. Close to our approach are the Dynamical Renormalization Group
 approach by Boyanovsky and collaborators~\cite{bv,bvhs}, and~Millington's and Pilaftsis' formulation~\cite{Millington:2012pf,Millington:2013qpa} of
 non-equilibrium thermal field theory. Specific to our approach is the use of the
 retarded-advanced ($R$-$A$) basis, where the Keldysh ($K$) propagator ($D_K$) is also separated into its advanced and retarded parts ($D_{K,A}$ and $D_{K,R} \, ,$ respectively). 
 The formalism~\cite{Dadic:1999bp,Dadic:2002wv,Dadic:2009zz,Dadic:2019lmm}
 is equivalent to the approach of Boyanovsky and De Vega~\cite{bv};
 nevertheless, in~the evaluation of production of direct photons,
 the difference is that we are calculating in the energy-momentum
 representation and~avoid early approximations and~simplifications.

\begin{figure}[H]

\centering

\includegraphics[width=12 cm]{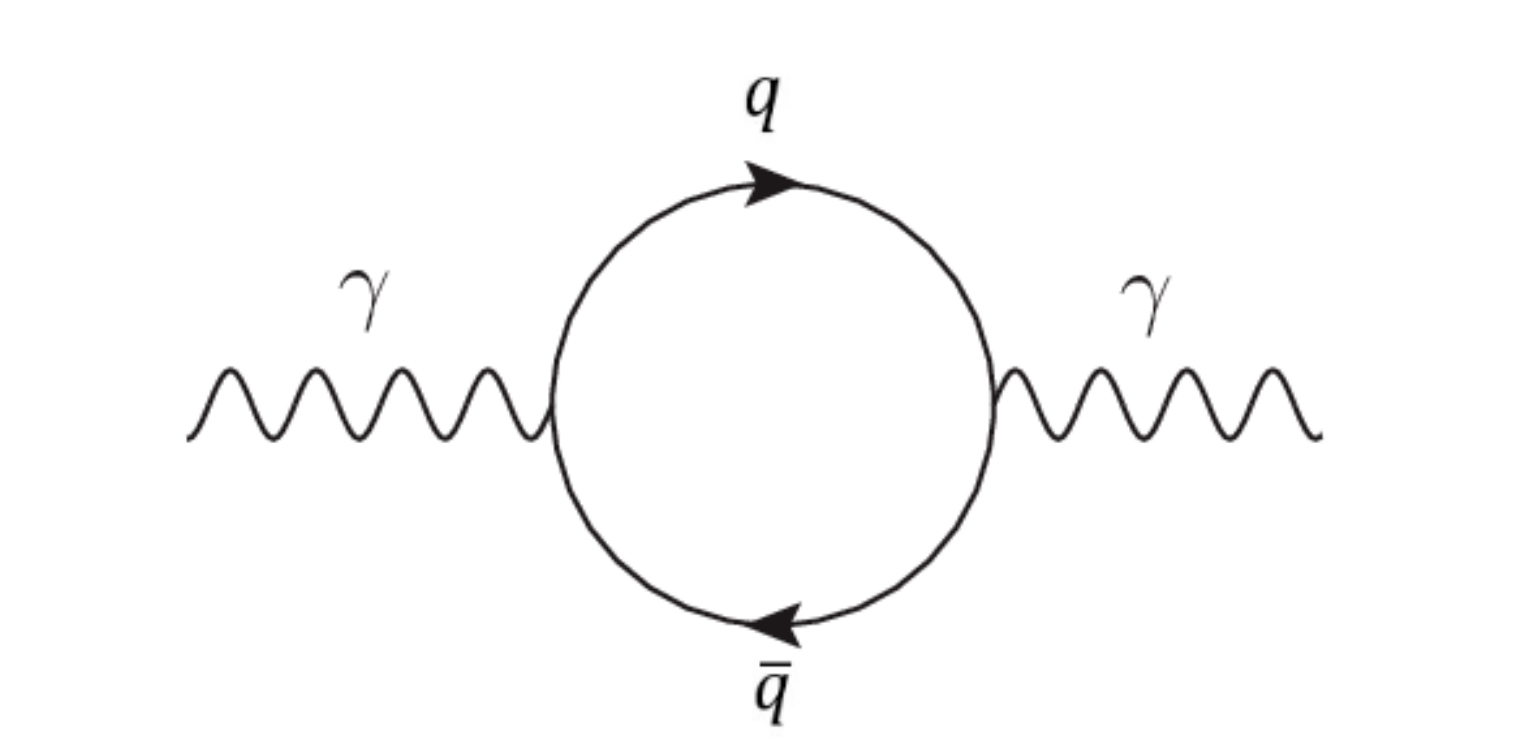}

\caption{The lowest-order correction to the vacuum polarization
 $\Pi^{\rho\sigma}$ of photons, due to quarks of the flavor $q$.
\label{vacuumPolarizationQEDqqbar}}

\end{figure}

Only having perturbative QED interactions, photons are ``clean'' probes of the quark-gluon
plasma (QGP), regardless of whether it is in the regime of nonperturbative or perturbative
QCD. For~temperatures that are not much higher than the (pseudo)critical temperature $T_c$
of the crossover transition, QGP is still strongly coupled for sure~\cite{Baym:2016wox,Pasechnik:2016wkt}. However, there are claims~\cite{Paquet:2015lta}
that, in photon production calculations, one can rely on perturbative QCD corrections. These
claims are supported by lattice around and even below $T \sim 1.3 T_c$~\cite{Ghiglieri:2016tvj}.
Even in references that stress the nonperturbative character of QCD significantly
above $T_c$, QGP begins approaching its perturbative regime beyond $T \sim 2 T_c$~\cite{Ding:2015ona},
i.e.,~above (250--300) MeV. Moreover, observables that are dominated at high $T$
by quark rather than gluon contributions seem to approach perturbative behavior
earlier~\cite{Ding:2015ona}, and~the fermionic sector is weakly coupled for 
$T > 300$ MeV~\cite{Haque:2014rua}. Thus, in~the very early, pre-equilibrium phase,
where temperature $T$ is not even defined, but~the average energy per particle is
high enough, i.e.,~higher than $k_B \times 300$ MeV, one can assume that the
asymptotic QCD regime is reached, and~asymptotic freedom makes quarks move around quasi-free.
Subsequently, in~the lowest approximation, one can neglect the dressing of quark-photon
vertices by QCD interactions. In~particular, the~fully dressed quark-photon vertex
in the vacuum polarization diagram of photons is replaced by the free quark-photon
 vertex $\, e_{\rm q}\,\gamma_\mu \,$, as~in Figure~\ref{vacuumPolarizationQEDqqbar}.

In any case, in~the high-energy phase, various flavors of quarks and antiquarks will be present and the~energy of the heavy-ion collisions determine the number of active quark flavors.
 With expansion and cooling, the average energy per particle drops continuously until
 the phase in which quarks became confined, and~the system turns into hadron matter,
 where strong interactions dominate. Finally further expansion produces
 interparticle distances bigger than the range of strong interactions,
 and~ the system decays to individual~particles.
 
 One expects the production of highest energy photons in the early stage of HIC.
 The mechanism that is discussed in this paper only produces photons in this~stage.
 
  The photons, which do not interact strongly with quarks and gluons,
  escape relatively easily, carrying the valuable information on  early stage of HIC.
  In this stage, the~uncertainty relations allow large energy uncertainty producing fast~oscillations.

The Dyson--Schwinger equation for photon $D_{K,R}$ propagator requires renormalization~\cite{Giambagi:1972,Thooft:1972,Ashmore:1972,Cicuta:1972,Wilson:1973,Kislinger:1976,Donoghue:1983,Johansson:1986,Kobes:1989,Keil:1989,LeBellac:1990,Elmfors:1992,Eijck:1996,Chapman:1997,Nakkagawa:1997,Baacke:1998,Esposito:1998,Knoll:2002,Jakovac:2005,Arrizabalaga:2007,Blaizot:2007,Blaizot:2015,RyderBook}
of divergent $\Pi_{R}$ and $\Pi_{A}$ vacuum polarizations.
 The related problems emerging are an additional energy-not-conserving vertex and
 regularized $\Pi_R$ not vanishing as $|p_0|\rightarrow\infty$ (potentially breaking
 causality). They are solved in full analogy to the case of out-of-equilibrium
 $\lambda \phi^3$ field theory, as~discussed in~\cite{Dadic:2019lmm}.
 The solution involves: energy integrations performed, while $d<4$,
 subtractions in $\Pi_F$, and~``reparation'' of causality in the products,
 like $D_R\Pi_R$ or $\Pi_RD_R$ and analogously for $\Pi_A$.

Our final result is finite.  In~particular, the~contributions that contain initial
 distribution functions of quarks and/or antiquarks are finite.  
  The calculation is straightforward, but~one should notice how potentially
  pinching term turns into the contribution linear in time $t$.

Prospects for further development are~discussed.

\unskip


\section{Direct Photon~Production}

At the time $t$, number density of detected photons  of the polarization $e$
 $(e=e_0,e_1,e_2,e_3)$ and momentum $\vec p$ is: 
\begin{eqnarray}\label{NPf}
\langle \, N_{\vec p,e,t} \, \rangle \, = \, (2\pi)^3{d{\cal N}_e\over d^3x \, d^3p } \, ,
\end{eqnarray}
where $d{\cal N}_e$ is the number of photons  of the polarization $e$ inside the differential volumes $d^3x$ and $d^3p$ of the coordinate and momentum~spaces.

The number density is connected to the equal time limit of the $K$ (Keldysh) component of
 the dressed photon propagator ${\cal D}_{\mu\nu}$ (for simplicity, we use Feynman gauge):
\begin{eqnarray}\label{2NPfi+1}
2 \langle \, N_{\vec p,e,t } \, \rangle + 1 \, = \,
 - \, {\omega_{p}\over 2\pi}\lim_{t_1\rightarrow t}\int dp_0{\cal D}_{K,t_1,t,\mu\nu}(p)e^{\mu}e^{*\nu} \, .
\end{eqnarray}

There are various definitions~\cite{Blaizot:2001nr,wbvl,Garbrecht:2002pd} of number density based on the product of creation and annihilation operators, $a^\dagger$ and $a$. They are mostly equivalent,
 at least at low orders. Our definition is adapted to the presently used formalism, as~the dressed
 Keldysh propagator ${\cal D}_{K} = {\cal D}_{K,R} - {\cal D}_{K,A}$ is easily calculated from
 perturbation expansion for matrix propagators (${\cal D}_{ij}$) as well as from Dyson--Schwinger~equation.

As expected, at~the lowest order of the perturbation expansion, it is 
\begin{eqnarray}\label{NP0}
2 \langle \, N_{\vec p,e,t }^0 \, \rangle + 1 \, = \, 2\, f_e(\vec p) \, + \, 1 \, ,
\end{eqnarray}
where the function $f_e(\vec p)$ is the initial ($t=0$) distribution function for photons.
However, they freely escape from the medium after they are created in the collision, and~
do not accumulate in the quark medium created by the collision of nucleons. Thus, we set
 $f_e(\vec p) = 0$ as the initial condition. We are only interested in the photons from the early quark~phase.


The initial distributions of quarks ($n_+(\vec p)$) and of antiquarks ($n_-(\vec p)$)
are the input functions for out-of-equilibrium field theory, where they are independent.
All of the initial distribution functions ($f_{\rm i} = f_e, n_\pm$) have to satisfy
 the following conditions: $\int \, d^3p f_{\rm i}(\vec p) < \infty $,
 $\int d^3p \, \omega_{\vec p} \, f_{\rm i}(\vec p)<\infty $, 
 saying, respectively, that the probability and the average energy are finite.
 There is no condition regarding~analyticity.

In principle, highly anisotropic situations can be considered. Nevertheless, in~many applications, the~basic assumption is that the system is close to equilibrium and,~in
such cases, a~natural choice for $n_\pm$ is the Fermi--Dirac distribution function, which is isotropic:
\begin{equation}
\label{Fermi-Dirac}
n_\pm(\vec p) \, \to \, n_\pm(\omega_p) \, = \,
 {\, 1 \over e^{(\omega_p - \mu_\pm)/T} + 1 \, }~.
\end{equation}

This was the choice of Wang and Boyanovsky~\cite{bv}, and~we will also use it when
we compare our result to~theirs.

 Even for out-of-equilibrium situations, some phenomenologists speak loosely
 about temperature $T$ and chemical potential $\mu$ as the characterization of average single particle energy and particle number. We expect that, by fitting the measured direct photon data,
 one can extract some knowledge regarding $f(\vec p)$ function.

In comparison with the cross-section, which is proportional to time derivative of exclusive number
of particles, the number of ``detected'' photons is inclusive. It is the density number of photons
  (i.e.,~``gain minus loss'') detected until the time $t$ (yield).

In order to calculate the first nontrivial order contributions to $\langle \, N_{\vec p,e,t} \, \rangle$
in (\ref{2NPfi+1}), for~${\cal D}_{K,t_1,t_2,\mu\nu}(p)$, we use Dyson--Schwinger Equations~(\ref{DSDP}),
where time and momenta variables are suppressed  for compactness,
and where $*$ denotes the convolution product defined in Appendix \ref{a1} by Equation~(\ref{pgf}):
\begin{equation}
\begin{array}{l}
{\cal D}_{\mu\nu,K} \, = \, { D}_{\mu\nu,K}
+i[D_{\mu\rho,R}*\Pi^{\rho\sigma}_{K}*{ D}_{\sigma\nu,A}
+D_{\mu\rho,R}*\Pi^{\rho\sigma}_{R}*{ D}_{\sigma\nu,K}
+D_{\mu\rho,K}*\Pi^{\rho\sigma}_{A}*{ D}_{\sigma\nu,A}] \,\, ,
\cr
\\
\Pi^{\rho\sigma}_{K} \, = \, -\Pi^{\rho\sigma}_{K,R}+\Pi^{\rho\sigma}_{K,A} \,\, ,
\end{array}\label{DSDP}\end{equation}
where all vacuum polarizations  $\Pi^{\rho\sigma}$ are of the lowest order (one loop), as in Figure~\ref{vacuumPolarizationQEDqqbar}.


%
The expression (\ref{DSDP}) contains two types of problems: divergences~\cite{Dadic:2019lmm}
 and vertices~\cite{Dadic:1998yd,Dadic:1999bp,Dadic:2002wv,Dadic:2009zz}.

The divergences are contained in $\Pi^{\rho\sigma}_{R}$ and $\Pi^{\rho\sigma}_{A}$. These
divergences are descendants of the divergence of the Feynman vacuum polarization $\Pi^{\rho\sigma}_{F}$.
In the dimensional regularization at $d<4$, they are regulated (i.e.,~finite) and can be subtracted. Other vacuum polarizations $\Pi^{\rho\sigma}_{K}$ are expected to be~finite.

Vertices are of the following three types:

\begin{enumerate}
\item Vertices  with at least one outgoing retarded propagator or incoming advanced propagator.
Energy conservation is achieved by simple integration over the energy of such propagator, by closing the contour from above for the retarded and from below for the advanced propagator.
This case includes the cases when there are more than one such propagators connecting the same vertex. In~this case, the loop integrals should not diverge.
  The convolution product  $*$ containing them turns
 into the usual algebraic product. Examples: the vertex
between the two-point functions $D_{\mu\rho,R}$ and  $\Pi^{\rho\sigma}_{K,R}$;
between $\Pi^{\rho\sigma}_{K,A}$ and ${ D}_{\sigma\nu,A}$;
between  $\Pi^{\rho\sigma}_{R}$ and $D_{\mu\rho,K,R}$;
between ${ D}_{\sigma\nu,K,A}$ and  $\Pi^{\rho\sigma}_{A}$,
as they appear in the Dyson--Schwinger equation terms
$D_{\mu\rho,R}*\Pi^{\rho\sigma}_{R}*{ D}_{\sigma\nu,K}
+D_{\mu\rho,K}*\Pi^{\rho\sigma}_{A}*{ D}_{\sigma\nu,A}$.

\item {Vertices without any outgoing retarded propagator or incoming advanced propagator.}
  They are lower in time than neighboring vertices.
 Closing the integration path always catches some singularities of the propagators.
These terms will not conserve energy, but~they oscillate in time, with~high frequency.
The examples are: the vertex between the two-point functions
 $D_{\mu\rho,R}$ and  $\Pi^{\rho\sigma}_{K,A}$;
between $\Pi^{\rho\sigma}_{K,R}$ and ${ D}_{\sigma\nu,A}$;
between  $\Pi^{\rho\sigma}_{R}$ and $D_{\mu\rho,K,A}$~; and,
between ${ D}_{\sigma\nu,K,R}$ and  $\Pi^{\rho\sigma}_{A}$.

\item Vertices with at least one outgoing retarded propagator, or~at least one incoming advanced propagator, but~with two or more such propagators entering the same vertex, where the corresponding loop integral diverges.
These vertices should conserve energy, but~divergent integrals
make them ill-defined. At~$d<4$, the~loop integrals are regulated, and~
the usual closing of the integration contour leads to energy conservation.
This make them group 1. vertices. Examples of such vertices are the ones
between the two-point functions $D_{\mu\rho,R}$ and  $\Pi^{\rho\sigma}_{R}$,
as well as $\Pi^{\rho\sigma}_{A}$ and ${ D}_{\sigma\nu,A}$.
 Additionally, in~the case that the mentioned vertex is connected with yet another one, where this connection satisfies the condition for the group 1 vertices, the mentioned vertex immediately belongs to the group~1.
\end{enumerate}

These properties lead to the simplified version of Dyson--Schwinger equation at the first order:
\begin{equation}
\begin{array}{l}
{\cal D}_{\mu\nu,K} \, = \, { D}_{\mu\nu,K}
+i \, [ \, -D_{\mu\rho,R}\Pi^{\rho\sigma}_{K,R}*{ D}_{\sigma\nu,A}
+D_{\mu\rho,R}*\Pi^{\rho\sigma}_{K,A}{ D}_{\sigma\nu,A}
\cr\\
\qquad \qquad -D_{\mu\rho,R}\Pi^{\rho\sigma}_{R}*{ D}_{\sigma\nu,K,A}
+D_{\mu\rho,K,R}*\Pi^{\rho\sigma}_{A}{ D}_{\sigma\nu,A}
\cr\\
\qquad \qquad +D_{\mu\rho,R}\Pi^{\rho\sigma}_{R}{ D}_{\sigma\nu,K,R}
-D_{\mu\rho,K,A}\Pi^{\rho\sigma}_{A}{ D}_{\sigma\nu,A} \, ].
\end{array}\label{DSDPS}\end{equation}

Inserting this ${\cal D}_{\mu\nu,K}$ in Equation~(\ref{2NPfi+1})
and using the convolution product defined in Appendix \ref{a1}, reveals
that the one-loop contribution to the photon number density~is
\begin{equation}
\begin{array}{l}
\langle \, N^{1}_{\vec p,t } \, \rangle  \, = \, 
\sum_{\mu,\nu}g^{\mu\nu}{\omega_{p}\over 4\pi}
i\int dp_{01}dp_{02}dp_{0}
P_{t}(p_0,{p_{01}+p_{02}\over 2}){i\over 2\pi}{e^{-it(p_{01}-p_{02}+i\epsilon)} -1
\over p_{01}-p_{02}+i\epsilon}
\cr\\
\times[\, -D_{\mu\rho,R}(p_{01},\vec p)\Pi^{\rho\sigma}_{K,R}(p_{01},\vec p)
{ D}_{\sigma\nu,A}(p_{02},\vec p)
-D_{\mu\rho,R}(p_{01},\vec p)\Pi^{\rho\sigma}_{R}(p_{01},\vec p)
{ D}_{\sigma\nu,K,A}(p_{02},\vec p)
\cr\\
\quad +D_{\mu\rho,R}(p_{01},\vec p)\Pi^{\rho\sigma}_{K,A}(p_{02},\vec p)
{ D}_{\sigma\nu,A}(p_{02},\vec p)
+D_{\mu\rho,K,R}(p_{01},\vec p)\Pi^{\rho\sigma}_{A}(p_{02},\vec p)
{ D}_{\sigma\nu,A}(p_{02},\vec p)
\cr\\
\quad +D_{\mu\rho,R}(p_{01},\vec p)\Pi^{\rho\sigma}_{R}(p_{01},\vec p)
D_{\sigma\nu,K,R}(p_{02},\vec p)
-D_{\mu\rho,K,A}(p_{01},\vec p)\Pi^{\rho\sigma}_{A}(p_{02},\vec p)
D_{\sigma\nu,A}(p_{02},\vec p) \, ]~.
\end{array}\label{DSDP2}\end{equation}

The last two terms ($RRR$ and $AAA$) vanish in the limit of equal time.
The function $P_{t}(p_0,{p_{01}+p_{02}\over 2})$ is the projecting function
defined in the Appendix~\ref{a1}.

In the rest of the text, we suppress the Lorentz~indices.

Next, we perform equal time limit procedure.
Equal time limit of the product of two or more retarded functions vanishes,
as one can, in~that case, close the integration contour $dp_0$  in a way to
catch no singularity. Such a product of advanced functions also vanishes.
 The diverging $\Pi_R$ and $\Pi_A$ are represented by vacuum polarizations
 in the matrix representation~\cite{Dadic:2019lmm} as
\begin{equation}
\begin{array}{l}
\Pi_{j,k}=-\Pi_{K,R}+\Pi_{K,A}
- \, k \, \Pi_{R}- \, j \, \Pi_{A}   \qquad (i,j = 1,2)    \,\, ,
\cr\\
\Pi_{11}=\Pi_{F}
-\Pi_{n_+,n_-,R}+\Pi_{n_+,n_-,A}  \,\, , 
\cr\\
Re\Pi_{R(A),n_\pm=0}(p)=Re\Pi_{F} ~,
\end{array}\label{g2KRR}\end{equation}
where $\Pi_{n_+,n_-,R(A)}$ are the corrections to $\Pi_{F}$ by finite $n_\pm$.

Calculated at $d<4$ and regularized by subtraction of constant term, the~closing
the integration path over $dp_{01}$ from above shows that the subtraction term~vanishes.

 Integration over $dp_0$ removes (see Appendix \ref{a1}) the
  $P_t(p_0,p'_0)$ projector.

In the first two terms in the square bracket in Equation~(\ref{DSDP2}), one
can close the $dp_{02}$ integration path from above and catch only the
singularities of $D_A$ and $D_{K,A}$. The~integration over $dp_{01}$
can be closed from below with the singularities caught from $D_R$ and
$D_{K,R}$ and from $\Pi_{R}$ and $\Pi_{K,R}$.
Important to notice is that the poles of both, $D_A$ and $D_{K,A}$, are equal
$[\bar p_{02}]_{1(2)}=\pm \omega_p+i\epsilon$, for~$D_R$ and $D_{K,R}$
they are $[\bar p_{01}]_{1(2)}=\pm \omega_p-i\epsilon$.
Singularities  of $\Pi_{R}$ and $\Pi_{K,R}$ may be~complicated.

In the terms with only poles of propagators, the~vertex factor requires a limiting procedure
\begin{eqnarray}\label{DSDP3}
\lim_{\epsilon\rightarrow 0}
[{e^{-it(\bar p_{01}-\bar p_{02}+i\epsilon)} -1
\over \bar p_{01}-\bar p_{02}+i\epsilon}]_{\bar p_{01}=
\bar p_{02}*=\pm \omega_p-i\epsilon}=-it.
\end{eqnarray}

The corresponding terms are growing linearly with~time.

Similar, but~complex conjugated, are the contributions of the last
two terms in Equation~(\ref{DSDP2}). Thus,~we obtain 

\begin{equation}
\begin{array}{l}
\int_{-\infty}^{\infty}dp_0 \, {\cal D}^1_{K,t}(p_0,\vec p)
\, = \, - \, {i \pi t\over 2\omega_p^2}\sum_{\lambda=\pm}
[ \, -\Pi_{K,R}(\lambda\omega_p,\vec p)
+\lambda \Pi_{R}(\lambda\omega_p,\vec p)(1+2f_e(\omega_p))
\cr\\
 \qquad \qquad \qquad \qquad \qquad \qquad \qquad \quad + \, \Pi_{K,A}(-\lambda\omega_p,\vec p)
-\lambda \Pi_{A}(-\lambda\omega_p,\vec p) \left(1+2f_e(\omega_p) \, \right) \, ]
\cr\\
  \qquad \qquad \qquad  - \, {\pi\over 4\omega_p^2}\sum_{\lambda=\pm}
{e^{-i2t\lambda\omega_p}-1\over \omega_p}
\, [ \, \lambda \Pi_{K,R}(\lambda\omega_p,\vec p)
- \Pi_{R}(\lambda\omega_p,\vec p)(1+2f_e(\omega_p))
\cr\\
\qquad \qquad \qquad \qquad \qquad \qquad \qquad + \, \lambda\Pi_{K,A}(-\lambda\omega_p,\vec p)
- \Pi_{A}(-\lambda\omega_p,\vec p)(1+2f_e(\omega_p))]
\cr\\
+ \,  {1\over \omega_p}\sum_{\lambda=\pm}{\cal P}\int_{cut} {dp_{0}
\over p^2}{e^{-it(p_{0}-\lambda\omega_p)} -1
\over p_{0}-\lambda\omega_p}
\, [\lambda Im\Pi_{K,R}(p_{0},\vec p)
+ Im\Pi_{R}(p_{0},\vec p)(1+2f_e(\omega_p)) \, ]
\cr\\
+{1\over \omega_p}\sum_{\lambda=\pm}{\cal P}\int _{cut}{dp_{0}\over p^2}{e^{it(p_{0}-\lambda\omega_p)} -1
\over p_{0}-\lambda\omega_p}
 \,[ \, \lambda Im\Pi_{K,A}(p_{0},\vec p) -Im\Pi_{A}(p_{0},\vec p)(1+2f_e(\omega_p))]~,
\end{array}\label{DSDP4}\end{equation}
where $\cal P$ denotes the principal value of the~integrals.

We use the symmetries of the one-loop vacuum polarizations to simplify the above expression
\begin{equation}
\begin{array}{l}
 \Pi_{K,R}(-|p_{0}|,\vec p)=-\Pi^{*}_{K.R}(|p_{0}|,\vec p)
=\Pi^{*}_{K,A}(- |p_{0}|,\vec p)=-\Pi_{K,A}( |p_{0}|,\vec p),
 \cr\\
\Pi_{R}(- |p_{0}|,\vec p)=\Pi^{*}_{R}(|p_{0}|,\vec p)
=\Pi^{*}_{A}(- |p_{0}|,\vec p) = \Pi_{A}(|p_{0}|,\vec p)~.
\end{array}\label{SESYMi}\end{equation}

Our final result for the one-loop contribution to the photon number density is
\begin{equation}
\begin{array}{l}
\langle \, N^1_{\vec p,t }  \, \rangle \, = \, {\omega_p\over 4\pi}
\int_{-\infty}^{\infty}dp_0  \, {\cal  D}^1_{K,t}(p_0,\vec p)  \, = \,
\cr\\
{ t\over 2\omega_p}
[- Im \Pi_{K,R}(\omega_p,\vec p)+Im \Pi_{R}(\omega_p,\vec p)(1+2f_e(\omega_p))]
\cr\\
-{1 \over 4\omega_p^2}(1+2f_e(\omega_p))[(1-\cos{2t\omega_p})
Re \Pi_{R}(\omega_p,\vec p)
+\sin {2t\omega_p}
 Im \Pi_{R}(\omega_p,\vec p)]
\cr\\
+{2\over \pi^2\omega_p} \, {\cal P}\int_{p_0>0,~cut} { dp_{0}\over (p_0^2-\omega_p^2)^2}
[ p_0(1-\cos tp_0\cos t\omega_p)-\omega_p\sin tp_0\sin t\omega_p]
\cr\\
\times(1+2f_e(\omega_p))Im\Pi_{R}(p_{0},\vec p)~.
\end{array}\label{DSDP4}\end{equation}

In Appendix \ref{c1}, we give the results on the vacuum polarizations that are needed in Equation~(\ref{DSDP4}).
To evaluate it numerically, we, of course, also need the early-time distribution functions. 
They are presently still unknown, but~we are presently considering several {\it Ans\"atze}
proposed in literature as physically motivated ``educated~guesses''.


\vspace{6pt}

\section{Discussion of the~Results}

Thus, we end up with four sorts of terms. One of them is just the initial particle
distribution, {i.e.}, Equation~(\ref{NP0}). As~we explain after Equation~(\ref{NP0}),
this ``zeroth order term'' is not contained in Equation~(\ref{DSDP4}). 

\vskip 2mm

\subsection{Energy Conserving Terms}

\vskip 1mm

The terms appearing in the first square bracket in (\ref{DSDP4}) are the terms from
propagator poles satisfying $\bar p_{01}=\bar p_{02}^*$. These terms grow linearly with time
(see Equation~(\ref{DSDP3})).
At large times, these terms would be dominating. They have the other desired properties:
\begin{enumerate}[leftmargin=8mm,labelsep=3mm]
\item[(1)] They conserve~energy. 

\item[(2)]\textls[-25]{They vanish for the distribution functions satisfying detailed balance principle.
Indeed, the~bracket from (\ref{RAK1}) in Appendix \ref{c1} is the defect of
 detailed balance in all of the channels:}
\begingroup\makeatletter\def\f@size{9}\check@mathfonts
\def\maketag@@@#1{\hbox{\m@th\normalsize\normalfont#1}}%
\begin{equation}
\begin{array}{l}
\Theta (p_0-q_0)\Theta (q_0)
[n_+(\omega_{p-q}) n_-(\omega_{q})(1+f(\omega_p)) 
(1-n_+(\omega_{p-q}))(1-n_-(\omega_{q}))f(\omega_p)]
\cr\\
+\Theta (q_0-p_0)\Theta (-q_0)
[n_-(\omega_{p-q}) n_+(\omega_{q})(1+f(\omega_p)) 
(1-n_-(\omega_{p-q}))(1-n_+(\omega_{q}))f(\omega_p)]
\cr\\
+\Theta (q_0-p_0)\Theta (q_0)
[(1-n_-(\omega_{p-q})) n_-(\omega_{q})(1+f(\omega_p)) 
-n_-(\omega_{p-q})(1-n_-(\omega_{q}))f(\omega_p)]
\cr\\
+\Theta (p_0-q_0)\Theta (-q_0)
[(1-n_+(\omega_{p-q}))n_+(\omega_{q})f(\omega_p)
\cr\\
-n_+(\omega_{p-q})(1- n_+(\omega_{q}))(1+f(\omega_p)) 
(1-n_+(\omega_{p-q}))n_+(\omega_{q})f(\omega_p)] \, ].
\end{array}\label{DDB}\end{equation}
\endgroup

In Equation~(\ref{DDB}), unpolarized photons are assumed; hence, no subscripts appear on their distribution functions $f(\omega_p)$.


\item[(3)] They are proportional to the lowest order Collision integral. Nevertheless, their contribution vanishes, owing to the kinematical limitations. (Otherwise, these terms would correspond to the contribution from the usual $S$-matrix formalism.)
\end{enumerate}

 \subsection{Term Containing $Re\Pi_R$}

\vskip 1mm

The second term in Equation~(\ref{DSDP4}) is
the term from propagator poles satisfying $\bar p_{01} = -\bar p_{02}^*$.
Energy is not conserved in this term.  The~contribution containing $Re\Pi_{R}$ requires
the renormalization of finite-time-path out-of-equilibrium $\phi^3$ QFT~\cite{Dadic:2019lmm}.
 The important points are: The~vertices in the products
$D_R * \Pi_{R}$ and $\Pi_{A} * D_A$ should conserve energy,
 but, owing to the divergences in $\Pi_{R}$ and $\Pi_{A}$, the~
 energy-delta function appears only if we integrate over intermediate
 energy while keeping $d<4$. Subsequently, thanks to properties of the
 convolution product~\cite{Dadic:1999bp}, we obtain
 $D_R*\Pi_{R}=D_R\Pi_{R}$ and $\Pi_{A}*{ D}_A=\Pi_{A}{ D}_A$.
 

The connection between vacuum polarizations~\cite{Dadic:2019lmm},
 together with symmetry relations (\ref{SESYMi}) gives for the
 vacuum parts $\Pi_{F}$ (where $n_\pm = 0$):
\begin{equation}
\begin{array}{l}
\Pi_{j,k}={1\over 2}[-\Pi_{K,R}+\Pi_{K,A}
-(-1)^k\Pi_{R}-(-1)^j\Pi_{A}]~,   \qquad   (j,k = 1,2)~,
\cr\\
Re\Pi_{R,n_\pm=0}=Re\Pi_{A,n_\pm=0}= Re\Pi_{F}~,
\end{array}\label{g2KRR}\end{equation}
where for the second line we have used the symmetries  (\ref{SESYMi}).
 Now, we obtain the renormalized value for
$Re\Pi_{ren,R,f=0}(q_0,\vec q)=Re\Pi_{R,f=0}(q_0,\vec q)-CT$, where $CT$ denotes the counter-term given by the $1/(4-d)$ term in Equation~(\ref{vap5}) or Equation~(\ref{vap6}) in Appendix~\ref{c1}. 

The quantity $Re\Pi_{ren,R,f=0}(q_0,\vec q)$, in~spite of having the label $R$, is
not a true retarded function, as it does not vanish when $|q_0|\rightarrow\infty $.
This creates a causality problem, which is repaired~\cite{Dadic:2019lmm} by considering
the composite objects $ D_R\Pi_{ren,R}$ and $\Pi_{ren,A} { D}_A$, which are retarded and advanced functions,~respectively.

\vskip 2mm

\subsection{Cut Contributions}

\vskip 1mm

These are the terms from propagator poles and singularities of vacuum
 polarizations (principal-value contributions in Equation~(\ref{DSDP4})).
 They are also~oscillating.

While, in the first term, we may identify kinetic energy
 $\bar p_0$ as $\pm \omega_p$,
 in second and third term it does not make sense as there are two different
 values for $\bar p_0$ for each~term.

At short times, all of the terms are important, as~they make sure that the
 uncertainty relations between energy and time are~satisfied.

\vskip 2mm

\subsection{Comparison to the Wang-Boyanovsky Result}

\vskip 1mm

The above expression should be compared to the Wang--Boyanovsky expression~\cite{wb}:
\begin{eqnarray}\label{dnb}
E \, {d{\cal N}(t)\over d^3p \, d^3x}={1\over 4\pi^4}\int_{-\infty}^{\infty}dp_0 
{1-\cos (p_0-E)t
\over (p_0-E)^2}R(p_0),
\end{eqnarray}
where $R(p_0)$ (in their notation, which is $f = n_+$ and ${\bar f} = n_-$
 in our notation (\ref{Fermi-Dirac})) is
\begin{equation}
\begin{array}{l}
R(p_0)={20\pi^2\alpha\over 3}\int {d^3k\over (2\pi)^3}
\cr\\
\left([1-(\hat p\hat k)(\hat p\hat q)][f(k)[1-\bar f(q)]
\delta(p_0-k+q)+[1-f(k)]\bar f(q)]
\delta(p_0+k-q)\right.
\cr\\\left.
+[1+(\hat p\hat k)(\hat p\hat q)]f(k)\bar f(q)\delta(p_0-k-q)
+[1-f(k)][1-\bar f(q)]\delta(p_0+k+q)\right),
\end{array}\label{sigb}\end{equation}
and where $k=|\vec k|$, $\hat k={\vec k \over k}$, $\hat p={\vec p \over E}$,
 $\vec q=\vec p-\vec k$, $q=|\vec q|$, and~$\hat q=\vec q/q$.

Despite a number of common features, there are significant differences:
Within the one-loop order approximation, our result is exact, while 
Refs.~\cite{wb,ng}   
ignore a few terms, and~make further approximations and~simplifications.

In our case, the emitted photons have undefined energy in accordance with
the Heisenberg uncertainty principle, similar to the findings of Millington and Pilaftsis in the context of a simple self-interacting theory of a real
 massive scalar field~\cite{Millington:2012pf}.
For Equation~(\ref{dnb}), Wang and Boyanovsky claim~\cite{wb} that the photons are on mass shell, i.e.,~$p_0=\omega_p$. As~the average of $\omega_p$
diverges~\cite{Arleo:2004gn}, it means infinite energy emitted in photons.
In our case, the~infinities are subtracted from $\Pi_{R(A)}$.
 It is evident that other terms are not ``dangerous''.

Our result allows for the distribution functions for $q$ and $\bar q$ to be determined
phenomenologically, i.e.,~not determined by $T,\mu$.

\section{\bf~Conclusions}

In this paper, we have calculated the production of photons from the early stage
 of quark gluon plasma, within~the finite-time-path out-of-equilibrium QFT.
The renormalization of mass divergence is analogously performed to the method that was developed
 for renormalization of $\lambda\phi^3$ out-of-equilibrium QFT~\cite{Dadic:2019lmm}.
The result is finite. 
The contribution in Equation~(\ref{DSDP4}) oscillates (similarly as in Ref.~\cite{Millington:2012pf})
in a way that is consistent with uncertainty relations between energy and time
 (except for energy conserving term, which vanishes for kinematical reasons). It is oscillating with the period $\propto 1/\omega_p$. Thus, after~a few periods it will be
 dominated by other, higher order terms, in~a lower energy-density medium, but still in pre-equilibrium.  Nevertheless, it is important, as this period may provide the highest energy photons. 

The result (\ref{DSDP4}) inspires further investigation that is necessary to compare and predict the production of photons by numerical calculations at the
 lowest order and develop methods in order to obtain higher order contributions.
We are in the process of performing the numerical analysis (to be published
 separately) aiming to investigate the following aspects:
\begin{enumerate}
\item Equation~(\ref{DSDP4}) contains renormalized $Re\Pi_R(\vec p)$ linearly, whereas in usual $S$-matrix calculations it appears quadratically in higher orders of
 the perturbation expansion. Thus, Equation~(\ref{DSDP4}). at least in principle, offers a possibility, albeit challenging, to~extract some information about
 $Re\Pi_R(\vec p)$ from~experiment.

\item One should distinguish the direct photon stage from the later stage in which the energy uncertainties are much smaller, but~higher order perturbation contributions become more important and even start to dominate (the damping phase).

\item The~early-time distributions of quarks ($n_+$) and antiquarks ($n_-$) are still unknown, and one should consider two very different situations: (a) the quarks are distributed isotropically
 and the probing functions could be taken as a thermalized Fermi--Dirac form like in Ref.~\cite{wb}.
 Or (b) the initial distribution of quarks may reflect the early stage distribution of nucleons.
 Some testing of {\it Ans\"atze} for the $n_\pm$ distributions will be necessary before reaching the final conclusion on the importance of the presented
 mechanism and its result (\ref{DSDP4}), but~we hope this will contribute to resolving the direct photon puzzle~\cite{David:2019wpt}.
\end{enumerate}

Other developments may follow in the more ambitious direction of renormalization
 of the full out-of-equilibrium QED (and QCD).


\vspace{6pt}
\authorcontributions{Conceptualization, I.D. and D.K. (Dubravko~Klabučar); Formal analysis, I.D.;
 Investigation, I.D., D.K. (Dubravko~Klabučar) and D.K. (Domagoj~Kuić); Methodology, I.D. and
 D.K. (Dubravko~Klabučar); Software, D.K. (Domagoj~Kuić); Supervision, I.D. and D.K. (Dubravko~Klabučar);
 Validation, I.D., D.K. (Dubravko~Klabučar) and D.K. (Domagoj~Kuić); Writing---original draft,
 I.D.; Writing---review \& editing, I.D., D.K. (Dubravko~Klabučar) and D.K. (Domagoj~Kuić). All authors have read and agreed to the published version of the manuscript.} 

\funding{Domagoj Kuić acknowledges partial support of the European Regional Development Fund---the
 Competitiveness and Cohesion Operational Programme: KK.01.1.1.06---RBI TWIN SIN, as~well as
the Croatian Science Foundation (HrZZ) Project No. IP-2016-06-3347.
Dubravko Klabučar thanks for partial support to COST Actions CA15213 THOR and CA16214~PHAROS. }

\acknowledgments{I. Dadić acknowledges useful discussions with R. Baier in the
 early stages of this~paper.}

\conflictsofinterest{The authors declare no conflict of~interest.} 


%
%
%
%
%
%
%
%
%
%
%
%



\appendixtitles{yes} 

\appendix

\section{Convolution~Product}\label{a1}

In the next step, we perform the convolution products and equal time~limit.

 The convolution product is defined as 
\begin{eqnarray}\label{pgf}
C=A*B \Leftrightarrow C(x,y)=\int_0^{\infty} dz A(x,z)B(z,y)~.
\end{eqnarray}

In terms of Wigner transforms of projected functions~\cite{Dadic:1999bp}, it~becomes
\begin{eqnarray}\label{ffffi}
C_{X_0}(p_0,\vec p)=\int dp_{01}dp_{02}
\, P_{X_0}(p_0,{p_{01}+p_{02}\over 2}) \,
{1\over 2\pi}{ie^{-iX_0(p_{01}-p_{02}+i\epsilon)} 
\over p_{01}-p_{02}+i\epsilon} \,
A_{\infty}(p_{01},\vec p) \, B_{\infty}(p_{02},\vec p)~,
\end{eqnarray}
where
\begin{eqnarray}\label{sftf}
P_{X_0}(p_0,p'_0)={1\over 2\pi}\Theta(X_0)\int_{-2X_0}^{2X_0}ds_0e^{is_0(p_0-p'_0)}
={1\over \pi}\Theta(X_0){\sin\left( 2X_0(p_0-p'_0)\right)\over p_0-p'_0},
\end{eqnarray}
and
\begin{eqnarray}\label{isftf}
e^{-is_0p'_0}\Theta(X_0)\Theta(2X_{0}+s_{0})\Theta(2X_{0}-s_{0})
=\int dp_0e^{-is_0p_0}P_{X_0}(p_0,p'_0).
\end{eqnarray}

 We have defined~\cite{Dadic:1999bp} the following properties:
(1) the function of $p_0$ is analytic
above (below) the real axis, (2) the function  goes to zero
as $|p_0|$ approaches infinity in the upper (lower) semi-plane.
The~choice above (below) and upper (lower) refers to Retarded (Advanced) functions.

In the following cases the product simplifies even further. These cases are:
1. If~$A$ is advanced function and $B$ advanced, retarded, or~even constant function.
2. If~$B$ is retarded function and $A$ advanced, retarded, or~even constant function.
Then the product becomes:
\begin{eqnarray}\label{fff4i}
C_{X_0}(p_0,\vec p)=\int dp_{01}P_{X_0}(p_0,p_{01})
A_{\infty}(p_{01},\vec p)B_{\infty}(p_{01},\vec p).
\end{eqnarray}

Note that when $X_0 \to \infty$, (\ref{sftf}) becomes the Dirac $\delta$-function,
and the convolution (\ref{fff4i}) reduces to the algebraic product:
$ \quad C_{\infty}(p_0,\vec p) \, = \, A_{\infty}(p_{0},\vec p) \, B_{\infty}(p_{0},\vec p)$.

However, we omit the subscript $_\infty$ throughout this paper.


\section{QED---Propagators}

The propagators in the covariant gauge and in the $R,A$ basis are:
\begin{equation}
\begin{array}{l}\label{pRAK}
D_{\mu\nu,R}(p)=[g_{\mu\nu}-(1-a)
{p_{\mu}p_{\nu}\over p^2+2ip_0\epsilon}]\Delta_{R}(p),
\cr\\
D_{\mu\nu,A}(p)=[g_{\mu\nu}-(1-a)
{p_{\mu}p_{\nu}\over p^2-2ip_0\epsilon}]\Delta_{A}(p)
\cr\\
\Delta_{R(A)}(p)={i\over p^2\pm 2ip_0\epsilon}~.
\end{array}\end{equation}

It would be interesting to obtain $D_{{\mu\nu},K}(p)$ as well as the number operator
 for different gauges. For~simplicity we set $a=1$, obtaining Feynman~gauge.

Initial densities for two transversal polarizations $e_{1,t}$, $e_{2,t}$ (linear or
circular, perpendicular to $(|\vec p|,\vec p)$ and mutually perpendicular) are given
 as $f_1(\vec p)$, $f_2(\vec p)$.  They could be joined by density $f_l(\vec p)$
 for  ``unphysical'  longitudinal polarization $e_l=(0,\vec p/|\vec p|)$  and density
 $f_0(\vec p)$ for timelike polarization $e_0=(1,0)$. As~the longitudinal and timelike
  densities do not evolve with time one can fix them in a various ways.
  In particular one can set them equal $f_l(\vec p)=f_0(\vec p)$, or~vanishing
 $f_l(\vec p)=f_0(\vec p)=0$. This is included in the definition of gauge.
The vectors defined above, form a new basis in four vector space:
$e_0, e_1,e_2,e_3=e_l$ and $g_{\mu,\nu}$ is easily transformed to this basis:

$g_{ij}=\sum_{\mu,\nu}g_{\mu\nu} \, e_i^{*\mu}e_j^{\nu}$,

$g_{\mu\nu}=\sum_{i,j}g_{ij} \, e^{*i}_{\mu}e^j_{\nu}$,

 $g_{00}=1$, $ g_{11}= g_{22}=g_{33}=-1$.

\vskip 1mm

Then:
\begin{equation}
\begin{array}{l}\label{pRAKF}
D_{\mu\nu,R}(p)=g_{\mu\nu} \, \Delta_{R}(p),
\cr\\
D_{\mu\nu,A}(p)=g_{\mu\nu} \, \Delta_{A}(p),
\cr\\
D_{{\mu\nu,K}}(p)
=2\pi\delta(p^2)\sum_{i,j}g_{ij}e^{*i}_{\mu}e^j_{\nu}[1+2f_{i}(\omega_p)]
\cr\\
=D_{{\mu\nu},K,R}(p)-D_{{\mu\nu},K,A}(p),
\cr\\
D_{{\mu\nu},K,R}(p)=-D_{{\mu\nu},K,A}(-p)
\cr\\
=-{p_0\over\omega_p}\sum_{i,j}g_{ij} \, e^{*i}_{\mu} \, e^j_{\nu}
[1+2f_{i}(\omega_p)] \, \Delta_{R}(p),
\cr\\
\omega_p=|\vec p|.
\end{array}\end{equation}

One needs spinor propagator (we omit  the label $_{\infty}$, it is understood
that whenever the time label is omitted it should be $_{\infty}$)
\begin{equation}
\begin{array}{l}\label{RAKS}
S_{R}(p) \, = \, (\pB+m)\, G_{R}(p,m)~,
\cr\\S_{A}(p) \, = \, (\pB+m)\, G_{A}(p,m)~,
\cr\\
G_{R(A)}(p,m) \, = \, {-i\over p^2-m^2\pm 2ip_0\epsilon}~,
\cr\\S_{K}(p)
=(\PB+{mp_0\over\omega_p}) \, 2\pi\delta(p^2-m^2) \, [1-2n(\omega_p)]=S_{K,R}(p)-S_{K,A}(p),
\cr\\
S_{K,R}(p)=-[1-2n(\omega_p)] \, (\PB+ {mp_0\over\omega_p}) \, G_{R}(p,m),
\cr\\
S_{K,A}(p)=-[1-2n(\omega_p)] \, (\PB+ {mp_0\over\omega_p}) \, G_{A}(p,m),
\cr\\
\pB=\gamma^{\mu}p_{\mu},~p=(p_0,\vec p),
~\PB=\gamma^{\mu}P_{\mu},~P=(\omega_p,{p_0\over \omega_p}\vec p),
~~
\omega_p=\sqrt{\vec p^2+m^2 \, }.
\end{array}\end{equation}

$n(\omega_p)$ is initial fermion distribution function. However, initial fermion and antifermion
distribution functions can in general be different. For~unequal distributions one has
%
%
\begin{eqnarray}\label{fne}
n(p_0,\omega_p,\vec p)=\Theta(p_0)n_+(\omega_p,\vec p)+\Theta(-p_0)n_-(\omega_p,-\vec p)~.
\end{eqnarray}

Now we decompose $K$-propagator into it`s retarded and advanced part
\begin{equation}
\begin{array}{l}\label{SRAf}
S_{K}(p,m) \, = \, 
S_{K,R}(p,m)-S_{K,A}(p,m)~,
\cr\\
S_{K,R}(p,m) \, = \, - \, G_{R}(p,m) \, L(p_0, \vec p)~,
\cr\\
S_{K,A}(p,m) \, = \, - \, G_{A}(p,m) \, L(p_0, \vec p)~,
\end{array}\end{equation}
where
\begin{equation}
\begin{array}{l}\label{L}
L(p_0, \vec p) \, = \, [1-2n_+(\omega_p,\vec p)]{p_0+\omega_p\over 2\omega_p}~2m\Lambda_+(\omega_p,\vec p)
\cr\\
+[1-2n_-(\omega_p,-\vec p)]{p_0-\omega_p\over 2\omega_p}~2m\Lambda_-(\omega_p,-\vec p)~,
\cr\\
\Lambda_{+}(\omega_p,\vec p)={\gamma_0\omega_p-\vec \gamma\vec p+m\over 2m}~,
\cr\\
\Lambda_{-}(\omega_p,\vec p)={-\gamma_0\omega_p+\vec \gamma\vec p+m\over 2m}~,
\end{array}\end{equation}
where the projectors $\Lambda_{\pm}$ satisfy
$\Lambda_{\pm}(\omega_p,\vec p)^2=\Lambda_{\pm}(\omega_p,\vec p)$, while
$\Lambda_{+}(\omega_p,\vec p)\Lambda_{-}(\omega_p,\vec p)=0$,
$\Lambda_{-}(\omega_p,\vec p)\Lambda_{+}(\omega_p,\vec p)=0$, and~
$\Lambda_{-}(\omega_p,\vec p)+\Lambda_{+}(\omega_p,\vec p)=1$.
%

The above result can be rewritten as:
\begin{equation}
\begin{array}{l}\label{RAKk}
S_{K,R(A)}(p)=[1-2{\bar n}(\omega_p)]{i(\PB+mp_0/\omega_p)
\over p^2-m^2\pm 2ip_0\epsilon}
-2n_{\Delta} (\omega_p)
{i(\pB+m)\over p^2-m^2\pm 2ip_0\epsilon},
\cr\\
\bar n (\omega_p)={n_+ (\omega_p)+ n _-(\omega_p)\over 2},       ~~
n _{\Delta}(\omega_p)={n_+ (\omega_p) -n _-(\omega_p)\over 2}~,   
\end{array}\end{equation}

We can also obtain $S_{11}$,
\begin{equation}
\begin{array}{lll}\label{RAK11}
S_{11}(p)&=&{1-2{\bar n}(\omega_p)\over 2}{i(\PB+mp_0/\omega_p)
\over p^2-m^2+ 2ip_0\epsilon}
+{1-2n_{\Delta} (\omega_p)\over 2}
{i(\pB+m)\over p^2-m^2+ 2ip_0\epsilon}
\cr\\
&-&{1-2{\bar n}(\omega_p)\over 2}{i(\PB+mp_0/\omega_p)
\over p^2-m^2-2ip_0\epsilon}
+{1+2n_{\Delta} (\omega_p)\over 2}
{i(\pB+m)\over p^2-m^2-2ip_0\epsilon}~.
\end{array}\end{equation}

These propagators satisfy the following properties under inversion:

$S_{R}(-p)=-\bar S_{A}(p),
~~
S_{K,R}(-p)=-\bar S_{K,A}(p)$, where $\bar S$ is the propagator of antifermion
(i.e.,~with~the the replacement of $n_+$ by $n_-$).

Then,
\begin{equation}
\begin{array}{l}\label{nupaf}
\langle \, 1-2N_{\pm,\pm \vec p}(t)  \, \rangle  \, = \, {\omega_p\over m\pi}\int dp_0 
{1\over 4}Tr_{\pm}[S_{K,R,t}(p)-S_{K,A,t}(p)]\gamma_0.
\cr
\\
={\omega_p\over m\pi}\int dp_0 
{1\over 4}Tr\Lambda_{\pm}(\omega_p,\pm \vec p)
[S_{K,R,t}(p)-S_{K,A,t}(p)]\gamma_0.
\end{array}\end{equation}
where the subscript ``+'' indicates that the trace $Tr$ is taken over the
fermion degrees of freedom and ``$-$'' over the  antifermion degrees of freedom.
When acting on momentum \& spin eigenstates of fermions, $|+,\vec p,s \, \rangle$,
and of antifermions, $|-,-\vec p,s \, \rangle $ (both normalized to
$\langle \, \pm, \pm \vec p,s|\pm, \pm \vec p,s \, \rangle \, = \, 1/m $),
the projectors $\Lambda_{\pm}$ satisfy

$\Lambda_{+}(\omega_p,\vec p)|+,\vec p,s  \, \rangle \, = \, |+,\vec p,s  \, \rangle \quad $
 and
$\quad  \Lambda_{-}(\omega_p,\vec p)|+,\vec p,s \, \rangle \,  = \, 0~, \quad\quad $ while

$\Lambda_{+}(\omega_p,\vec p)|-,-\vec p,s \, \rangle \, = \, 0 \quad $ and
$\quad  \Lambda_{-}(\omega_p,\vec p)|-,-\vec p,s \, \rangle \, = \, |-,-\vec p,s\, \rangle$.

%


\section{One-Loop Vacuum~Polarizations }\label{c1}


In this appendix, the~vacuum polarizations are calculated as contributions of a single
 quark flavor q with the charge $e_{\rm q} = C_{\rm q} \, e$, where $C_{\rm q}$ can
 take the values $C_{\rm q} = \pm 1/3,\pm 2/3$. ($e$ is the electron charge.) Each flavor
 has its initial distribution functions $n_{\pm}(\omega_p)$, and~mass $m_{\rm q}$.
 In the present paper, we used the mass symbol $m$ without the flavor subscript q
 because our analysis pertains to a single flavor.
 To~obtain the full result, one has to sum over all active quark~flavors.

One obtains vacuum polarizations easily as one knows the perturbation expansion for matrix propagators.
Here we use $\bar S $ as a symbol for anti-fermion propagator. It differs from the fermion propagator
 by  the fact that the roles of $n_+$ and $n_-$ are interchanged.
For all one-loop vacuum polarizations in Equation (\ref{RAK10}),
$\Pi^1 = \Pi_{A}, \Pi_{R}, \Pi_{K}$ (the label $\infty$ is omitted for simplicity after (\ref{RAK10})),
\begin{eqnarray}\label{RAK10}
\Pi_{t}(p_0,\vec p)=\int_{-\infty}^{\infty}dp'_0 P_t(p_0,p_0')\Pi_{\infty}(p_0',\vec p)~,
\end{eqnarray}
\begin{eqnarray}\label{RAK10a}
&&
\Pi_{A}(p_0,\vec p)={ig^2\over 4} \int {dq_0d^{d-1}q\over (2\pi)^{d}} 
[S_{A}(p_0-q_0,\vec p-\vec q,m)\bar S_{K}(q_0,\vec q,m)
\cr\nonumber\\&&
 +S_{K}(p_0-q_0,\vec p-\vec q,m) \bar S_{A}(q_0,\vec q,m)]~,
\cr\nonumber\\&&
\Pi_{R}(p_0,\vec p)=-{ig^2\over 4} \int {dq_0d^{d-1}q\over (2\pi)^{d}} 
[S_{R}(p_0-q_0,\vec p-\vec q,m)\bar S_{K}(q_0,\vec q,m)
\cr\nonumber\\&&
+ S_{K}(p_0-q_0,\vec p-\vec q,m) \bar S_{R}(q_0,\vec q,m)]~,
\cr\nonumber\\&&
Im~\Pi_{A}(p_0,\vec p)=
-{g^2\pi\over 4}\int{d^{d-1}q\over (2\pi)^{d}}
\int dq_0\delta (q_0^2-\omega_{q}^2)\delta ((p_0-q_0)^2-\omega_{p-q}^2)
\cr\nonumber\\&&
\times sign[q_0(p_0-q_0)]
[Tr\gamma^0(\pB-\qB+m) \gamma^0(\qB+m)]_{|q_0|=\omega_q,~|p_0-q_0|=\omega_{p-q}}
\cr\nonumber\\&&
[\Theta (p_0-q_0)(1-2n_+(\omega_{p-q}))-(1-\Theta (p_0-q_0))(1-2n_-(\omega_{p-q}))
\cr\nonumber\\&&
+\Theta (q_0)(1-2n_-(\omega_{q}))-(1-\Theta (q_0))(1-2n_+(\omega_{q}))]~,
\cr\nonumber\\&&
Im~\Pi_{A}(p_0,\vec p)=
-Im~\Pi_{R}(p_0,\vec p)=-Im~\Pi_{A}(-p_0,\vec p)=Im~\Pi_{R}(-p_0,\vec p)~,
\cr\nonumber\\&&
Re~\Pi_{A}(p_0,\vec p)=Re~\Pi_{R}(p_0,\vec p)~,
\cr\nonumber\\&&
\Pi_{K}(p_0,\vec p)
=-\Pi_{K,R}(p_0,\vec p)+\Pi_{K,A}(p_0,\vec p)~,
\cr\nonumber\\&&
\Pi_{K,R}(p_0,\vec p)=
\cr\nonumber\\&&
{-ig^2\over 2} \int {dq_0d^{d-1}q\over (2\pi)^{d}} 
[S_{K,R}(p_0-q_0,\vec p-\vec q,m)\bar S_{K,R}(q_0,\vec q,m)
\cr\nonumber\\&&
+ S_R(p_0-q_0,\vec p-\vec q,m)\bar S_{R}(q_0,\vec q,m)]~,
\cr\nonumber\\&&
\Pi_{K,A}(p_0,\vec p)=
\cr\nonumber\\&&
{ig^2\over 2} \int {dq_0d^{d-1}q\over (2\pi)^{d}} 
[S_{K,A}(p_0-q_0,\vec p-\vec q,m)\bar S_{K,A}(q_0,\vec q,m)
\cr\nonumber\\&&
+S_{A}(p_0-q_0,\vec p-\vec q,m)\bar S_{A}(q_0,\vec q,m)]
\cr\nonumber\\&&
Re~\Pi_{K,A}(p_0,\vec p)=Re~\Pi_{K,R}(p_0,\vec p),
\cr\nonumber\\&&
Im~\Pi_{K,A}(p_0,\vec p)
\cr\nonumber\\&&
=-{g^2\pi\over 4}\int{d^{d-1}q\over (2\pi)^{d-1}}
\int dq_0\delta (q_0^2-\omega_{q}^2)\delta ((p_0-q_0)^2-\omega_{p-q}^2)
\cr\nonumber\\&&
\times sign[q_0(p_0-q_0)]
[Tr\gamma^0(\pB-\qB+m) \gamma^0(\qB+m)]_{|q_0|=\omega_q,~|p_0-q_0|=\omega_{p-q}}
\cr\nonumber\\&&
[[\Theta (p_0-q_0)(1-2n_+(\omega_{p-q}))-(1-\Theta (p_0-q_0))(1-2n_-(\omega_{p-q}))]
\cr\nonumber\\&&
\times
[\Theta (q_0)(1-2n_-(\omega_{q}))-(1-\Theta (q_0))(1-2n_+(\omega_{q}))]+1]~,
\cr\nonumber\\&&
Im~\Pi_{K,A}(p_0,\vec p)
=-Im~\Pi_{K,R}(p_0,\vec p),
\cr\nonumber\\&&
\omega_{q}^2=m^2+\vec q^2,~~~
\omega_{p-q}^2=m^2+(\vec p-\vec q)^2
\cr\nonumber\\&&
\delta (q_0^2-\omega_{q}^2)\delta ((p_0-q_0)^2-\omega_{p-q}^2)
\cr\nonumber\\&&
=\sum_{\lambda,\lambda'=\pm}{\delta (q_0-\lambda\omega_{q})\delta (p_0-q_0-\lambda'\omega_{p-q})
\over 4\omega_{q}\omega_{p-q}}~,
\end{eqnarray}
\begin{eqnarray}\label{RAK1}
&&
   Im \Pi_{K,R}(\omega_p,\vec p)+Im \Pi_{R}(\omega_p,\vec p)
   (1+2f(\omega_p))
 \cr\nonumber\\&&
 ={g^2\pi\over 4}\int{d^{d-1}q\over (2\pi)^{d-1}}
\int dq_0\delta (q_0^2-\omega_{q}^2)\delta ((p_0-q_0)^2-\omega_{p-q}^2)
\cr\nonumber\\&&
\times sign[q_0(p_0-q_0)]
[Tr\gamma^0(\pB-\qB+m) \gamma^0(\qB+m)]_{|q_0|=\omega_q,~|p_0-q_0|=\omega_{p-q}}
\cr\nonumber\\&&
\left[ \, \{ \, [\,\Theta (p_0-q_0)(1-2n_+(\omega_{p-q}))-(1-\Theta (p_0-q_0))(1-2n_-(\omega_{p-q}))]\right.
\cr\nonumber\\&&
\times[\, \Theta (q_0)(1-2n_-(\omega_{q}))-(1-\Theta (q_0))(1-2n_+(\omega_{q}))]+1 \, \}
\cr\nonumber\\&&
-[\, \Theta (p_0-q_0)(1-2n_+(\omega_{p-q}))-(1-\Theta (p_0-q_0))(1-2n_-(\omega_{p-q}))
\cr\nonumber\\&&
+ \, \Theta (q_0)(1-2n_-(\omega_{q}))-(1-\Theta (q_0))(1-2n_+(\omega_{q}))]
(1+2f(\omega_p)) \, ] \, .
  \end{eqnarray}

At $|q_0|=\omega_{q}$ and $|p_0-q_0|=\omega_{p-q}$,
 the bracket containing distribution functions
vanishes when the functions satisfy detailed balance condition in all channels:
\begin{equation}
\begin{array}{l}\label{DDD}
  Im \Pi_{K,R}(\omega_p,\vec p)+Im \Pi_{R}(\omega_p,\vec p)
   (1+2f(\omega_p))
 \cr\\
 ={g^2\pi\over 4}\int{d^{d-1}q\over (2\pi)^{d-1}}
\int dq_0\delta (q_0^2-\omega_{q}^2)\delta ((p_0-q_0)^2-\omega_{p-q}^2)
\cr\\
\times sign[q_0(p_0-q_0)]
[Tr\gamma^0(\pB-\qB+m) \gamma^0(\qB+m)]_{|q_0|=\omega_q,~|p_0-q_0|=\omega_{p-q}}
\cr\\
\times[ 4\Theta (p_0-q_0)\Theta (q_0)
[n_+(\omega_{p-q}) n_-(\omega_{q})(1+f(\omega_p)) 
(1-n_+(\omega_{p-q}))(1-n_-(\omega_{q}))f(\omega_p)]
\cr\\
+4\Theta (q_0-p_0)\Theta (-q_0)
[n_-(\omega_{p-q}) n_+(\omega_{q})(1+f(\omega_p)) 
(1-n_-(\omega_{p-q}))(1-n_+(\omega_{q}))f(\omega_p)]
\cr\\
+4\Theta (q_0-p_0)\Theta (q_0)
[(1-n_-(\omega_{p-q})) n_-(\omega_{q})(1+f(\omega_p)) 
-n_-(\omega_{p-q})(1-n_-(\omega_{q}))f(\omega_p)]
\cr\\
+4\Theta (p_0-q_0)\Theta (-q_0)
[(1-n_+(\omega_{p-q}))n_+(\omega_{q})f(\omega_p)
\cr\\
-n_+(\omega_{p-q})(1- n_+(\omega_{q}))(1+f(\omega_p)) 
(1-n_+(\omega_{p-q}))n_+(\omega_{q})f(\omega_p)]].
\end{array}\end{equation}

Notice that in this section of appendix, we have assumed that the initial photon distribution
  does not depend on the photon polarization $e$. Thus, we wrote $f(\omega_p)$ instead of $f_e(\omega_p)$.


\vskip 2mm

\underline{Regularized $\Pi_F$}

\vskip 1mm

For the dimensionally regularized $ \Pi_{F}$,   $\kappa \equiv 4-d > 0$, the~result is causal,
as  $|q_0|\rightarrow \infty$ implies $|\Pi_{\mu\nu,F,d}(q)|\rightarrow 0$.
The Feynman component ($n_\pm = 0$) of the vacuum polarization, expanded around small $\kappa$,
can be written as~\cite{RyderBook}
\begin{eqnarray}\label{vap5}
\Pi_{\mu\nu,F}(q)={e^2\over 2\pi^2} \,
(q_{\mu}q_{\nu}-q^2g_{\mu\nu}) \, 
[\, {1\over 3\kappa} \, - \, {\gamma_{\rm E}\over 6}
-\int_0^1dz \, z(1-z) \, \ln{q^2z(1-z) - m^2 \over 4\pi\mu^2}]~.
\end{eqnarray}

For small $q^2$~\cite{Dadic:2019lmm}, this becomes
\begin{eqnarray}\label{vap6}
&&\Pi_{\mu\nu,F}(q)={e^2\over 6\pi^2}
(q_{\mu}q_{\nu}-q^2g_{\mu\nu})({1\over \kappa}+{q^2\over 10m^2}+ ...)
\cr
\nonumber\\&&
={e^2\over 6\pi^2}
(q_{\mu}q_{\nu}-q^2g_{\mu\nu}){1\over \kappa} \, + \, {\rm finite}~.
\end{eqnarray}

 We see that the finite part of $\Pi_{\mu\nu,F}(q)$ is not vanishing
  when $|q_0|\rightarrow \infty$. This implies that also ``retarded'' and
  ''advanced'' part of it do not satisfy this requirement.
  To repair causality, we turned to the composite operators pointed out
 below Equation~(\ref{g2KRR}), namely $ D_R\Pi_{ren,R}$ and $\Pi_{ren,A} { D}_A$
 which satisfy this requirement. ($D_R$ and $D_A$ yield the $1/\omega_p^2$
 suppression of the second square bracket in Equation~(\ref{DSDP4}).)



\reftitle{References}



\begin{thebibliography}{999}

\bibitem{Baym:2016wox}
Baym, G.
Ultrarelativistic heavy ion collisions: The first billion seconds.
\emph{Nucl.\ Phys.\ A} \textbf{2016}, {\emph{956}}, 1. [\href{http://dx.doi.org/10.1016/j.nuclphysa.2016.03.007}{CrossRef}]

\bibitem{Pasechnik:2016wkt}
Pasechnik, R.; Šumbera, M.
Phenomenological Review on Quark---Gluon Plasma: Concepts vs. Observations.
\emph{Universe} {\bf 2017}, \emph{3}, 7. [\href{http://dx.doi.org/10.3390/universe3010007}{CrossRef}]


\bibitem{David:2019wpt}
David, G.
Direct real photons in relativistic heavy ion collisions. \emph{Rept.\ Prog.\ Phys.\ }  {\bf 2020}, \emph{83}, 046301. [\href{http://dx.doi.org/10.1088/1361-6633/ab6f57}{CrossRef}]

\bibitem{Schwinger:1960qe}
Schwinger, J.S. Brownian motion of a quantum oscillator. \emph{J.\ Math.\ Phys.\ }  {\bf 1961}, \emph{2}, 407. [\href{http://dx.doi.org/10.1063/1.1703727}{CrossRef}]

\bibitem{Keldysh:1964ud}
Keldysh, L.V. Diagram technique for nonequilibrium processes.
\emph{Sov.\ Phys.\ JETP} \textbf{1965}, {\emph 20}, 1018--1026.

\bibitem{KadanoffBook}
Kadanoff, L.P.; Baym, G.
\emph{Quantum Statistical Mechanics}; Benjamin: New York, NY, USA, 1962.

\bibitem{Danielewicz:1982kk}
Danielewicz, P.
Quantum Theory of Nonequilibrium Processes. 1.
\emph{Ann. Phys.} {\bf 1984}, \emph{152}, 239. [\href{http://dx.doi.org/10.1016/0003-4916(84)90092-7}{CrossRef}]

\bibitem{Chou:1984es}
Chou, K.C.; Su, Z.B.; Hao, B.L.; Yu, L.
Equilibrium and Nonequilibrium Formalisms Made Unified.
\emph{Phys.~Rept.\ }  {\bf 1985}, \emph{118}, 1. [\href{http://dx.doi.org/10.1016/0370-1573(85)90136-X}{CrossRef}]


\bibitem{Rammer:1986zz}
Rammer, J.; Smith, H.
Quantum field-theoretical methods in transport theory of metals.
\emph{Rev.\ Mod.\ Phys.\ }  {\bf 1986}, \emph{58}, 323. [\href{http://dx.doi.org/10.1103/RevModPhys.58.323}{CrossRef}]


\bibitem{Landsman:1986uw}
Landsman, N.P.; van Weert, C.G.
Real and Imaginary Time Field Theory at Finite Temperature and Density.
\emph{Phys.\ Rept.\ }  {\bf 1987}, \emph{145}, 141. [\href{http://dx.doi.org/10.1016/0370-1573(87)90121-9}{CrossRef}]


\bibitem{Calzetta:1986cq}
Calzetta, E.; Hu, B.L.
Nonequilibrium Quantum Fields: Closed Time Path Effective Action, Wigner Function and Boltzmann Equation.
\emph{Phys.\ Rev.\ D} {\bf 1988}, \emph{37}, 2878. [\href{http://dx.doi.org/10.1103/PhysRevD.37.2878}{CrossRef}]


\bibitem{Niemi:1987pm}
Niemi, A.J.
Nonequilibrium Quantum Field Theories. \emph{Phys.\ Lett.\ B} {\bf 1988}, \emph{203}, 425--432. [\href{http://dx.doi.org/10.1016/0370-2693(88)90196-7}{CrossRef}]


\bibitem{Remler:1990}
Remler, E.A.
Simulation of multiparticle scattering. \emph{Ann. Phys.} {\bf 1990}, \emph{202}, 351. [\href{http://dx.doi.org/10.1016/0003-4916(90)90229-H}{CrossRef}]


\bibitem{LeBellacBook}
Bellac, M.L. \emph{Thermal Field Theory}; Cambridge University Press: Cambridge, UK, {1996}.


\bibitem{Brown:1998zx}
Brown, D.A.; Danielewicz, P.
Partons in phase space.  \emph{Phys.\ Rev.\ D} {\bf 1998}, \emph{58}, 094003. [\href{http://dx.doi.org/10.1103/PhysRevD.58.094003}{CrossRef}]


\bibitem{Blaizot:2001nr}
Blaizot, J.P.; Iancu, E.
The Quark gluon plasma: Collective dynamics and hard thermal loops.
\emph{Phys.\ Rept.\ } {\bf 2002}, \emph{359}, 355--528. [\href{http://dx.doi.org/10.1016/S0370-1573(01)00061-8}{CrossRef}]


\bibitem{bv}
Boyanovsky, D.; de Vega, H.J.
Anomalous kinetics of hard charged particles: Dynamical renormalization group resummation.  \emph{Phys.\ Rev.\ D} {\bf 1999}, \emph{59}, 105019. [\href{http://dx.doi.org/10.1103/PhysRevD.59.105019}{CrossRef}]


\bibitem{bvhs}
Boyanovsky, D.; de Vega, H.J.; Holman, R.; Simionato, M.
Dynamical renormalization group resummation of finite temperature infrared divergences.
\emph{Phys.\ Rev.\ D} {\bf 1999}, \emph{60}, 065003.  [\href{http://dx.doi.org/10.1103/PhysRevD.60.065003}{CrossRef}]


\bibitem{devega}
Boyanovsky, D.; de Vega, H.J.; Wang, S.Y.
Dynamical renormalization group approach to quantum kinetics in scalar and gauge theories.
\emph{Phys.\ Rev.\ D} {\bf 2000}, \emph{61}, 065006. [\href{http://dx.doi.org/10.1103/PhysRevD.61.065006}{CrossRef}]


\bibitem{bvs0}
Boyanovsky, D.; de Vega, H.J.; Simionato, M.
Nonequilibrium quantum plasmas in scalar QED: Photon production, magnetic and Debye masses and conductivity.
\emph{Phys.\ Rev.\ D}  {\bf 2000}, \emph{61}, 085007. [\href{http://dx.doi.org/10.1103/PhysRevD.61.085007}{CrossRef}]


\bibitem{wbvl}
Wang, S.Y.; Boyanovsky, D.; de Vega, H.J.; Lee, D.S.
Real time nonequilibrium dynamics in hot QED plasmas: Dynamical renormalization group approach.
\emph{Phys.\ Rev.\ D} {\bf 2000}, \emph{62}, 105026. [\href{http://dx.doi.org/10.1103/PhysRevD.62.105026}{CrossRef}]


\bibitem{wb}
Wang, S.Y.; Boyanovsky, D.
Enhanced photon production from quark---gluon plasma: Finite lifetime effect.
\emph{Phys.\ Rev.\ D} {\bf 2001}, \emph{63}, 051702. [\href{http://dx.doi.org/10.1103/PhysRevD.63.051702}{CrossRef}]


\bibitem{ng}
Wang, S.Y.; Boyanovsky, D.; Ng, K.W.
Direct photons: A nonequilibrium signal of the expanding quark gluon plasma at RHIC energies.
\emph{Nucl.\ Phys.\ A} {\bf 2002}, \emph{699}, 819--846. [\href{http://dx.doi.org/10.1016/S0375-9474(01)01288-X}{CrossRef}]


\bibitem{bv68}
Boyanovsky, D.; de Vega, H.J.
Are direct photons a clean signal of a thermalized quark gluon plasma? \emph{Phys.~Rev.\ D} {\bf 2003}, \emph{68}, 065018. [\href{http://dx.doi.org/10.1103/PhysRevD.68.065018}{CrossRef}]


\bibitem{bvhep}
Boyanovsky, D.; de Vega, H.J.
Photon production from a thermalized quark gluon plasma: Quantum kinetics and nonperturbative aspects.
\emph{Nucl.\ Phys.\ A} {\bf 2005}, \emph{747}, 564--608. [\href{http://dx.doi.org/10.1016/j.nuclphysa.2004.10.006}{CrossRef}]


\bibitem{Arleo:2004gn}
Arleo, F.; Aurenche, P.; Bopp, F.W.; Dadic, I.; David, G.; Delagrange, H.; d’Enterria, D.G.; Eskola, K.J.
Hard~probes in heavy-ion collisions at the LHC: Photon physics in heavy ion collisions at the LHC. In \emph{CERN Yellow Book CERN-2004-009-D}; CERN: Geneva, Switzerland, 2004.


\bibitem{Millington:2012pf}
Millington, P.; Pilaftsis, A.
Perturbative nonequilibrium thermal field theory. \emph{Phys.\ Rev.\ D} {\bf 2013}, \emph{88}, 085009. [\href{http://dx.doi.org/10.1103/PhysRevD.88.085009}{CrossRef}]

\bibitem{Millington:2013qpa}
Millington, P.; Pilaftsis, A.
Thermal field theory to all orders in gradient expansion.
\emph{J.\ Phys.\ Conf.\ Ser.\ } {\bf 2013}, \emph{447}, 012071. [\href{http://dx.doi.org/10.1088/1742-6596/447/1/012071}{CrossRef}]


\bibitem{Dadic:1999bp}
Dadić, I.
Out-of-equilibrium thermal field theories: Finite time after switching on
the interaction: Fourier transforms of the projected functions.
\emph{Phys.\ Rev.\ D} {\bf 2000}, \emph{63}, 025011; Erratum in {\bf 2002}, \emph{66}, 069903. [\href{http://dx.doi.org/10.1103/PhysRevD.63.025011}{CrossRef}]


\bibitem{Dadic:2002wv}
Dadić, I.
Out-of-equilibrium TFT---energy nonconservation at vertices.
\emph{Nucl.\ Phys.\ A} {\bf 2002}, \emph{702}, 356. [\href{http://dx.doi.org/10.1016/S0375-9474(02)00724-8}{CrossRef}]


\bibitem{Dadic:2009zz}
Dadić, I.
Retarded propagator representation of out-of-equilibrium thermal field theories.
\emph{Nucl.\ Phys.\ A} {\bf 2009}, \emph{820}, 267C. [\href{http://dx.doi.org/10.1016/j.nuclphysa.2009.01.066}{CrossRef}]


\bibitem{Dadic:2019lmm}
Dadić, I.; Klabučar, D.
Causality and Renormalization in Finite-Time-Path Out-of-Equilibrium $\phi^3$ QFT.
\emph{Particles} {\bf 2019}, \emph{2}, 92--102. [\href{http://dx.doi.org/10.3390/particles2010008}{CrossRef}]


\bibitem{Paquet:2015lta}
Paquet, J.F.; Shen, C.; Denicol, G.S.; Luzum, M.; Schenke, B.; Jeon, S.; Gale, C.
Production of photons in relativistic heavy-ion collisions. \emph{Phys.\ Rev.\ C} {\bf 2016}, \emph{93}, 044906. [\href{http://dx.doi.org/10.1103/PhysRevC.93.044906}{CrossRef}]


\bibitem{Ghiglieri:2016tvj}
Ghiglieri, J.; Kaczmarek, O.; Laine, M.; Meyer, F.
Lattice constraints on the thermal photon rate. \emph{Phys.\ Rev.\ D} {\bf 2016}, \emph{94}, 016005. [\href{http://dx.doi.org/10.1103/PhysRevD.94.016005}{CrossRef}]
\bibitem{Ding:2015ona}
Ding, H.T.; Karsch, F.; Mukherjee, S.
Thermodynamics of strong-interaction matter from Lattice QCD.
\emph{Int.\ J.\ Mod.\ Phys.\ E} {\bf 2015}, \emph{24}, 1530007. [\href{http://dx.doi.org/10.1142/S0218301315300076}{CrossRef}]

\bibitem{Haque:2014rua}
Haque, N.; Bandyopadhyay, A.; Andersen, J.O.; Mustafa, M.G.; Strickland, M.; Su, N.
Three-loop HTLpt thermodynamics at finite temperature and chemical potential.
\emph{J.~High Energy~Phys.} {\bf 2014}, \emph{1450}, 027. [\href{http://dx.doi.org/10.1007/JHEP05(2014)027}{CrossRef}]


\bibitem{Giambagi:1972}
Bollini, C.G.; Giambiagi, J.J.
Dimensional Renormalization: The Number of Dimensions as a Regularizing Parameter.
\emph{Nuovo Cim.\ B} {\bf 1972}, \emph{12}, 20. [\href{http://dx.doi.org/10.1007/BF02895558}{CrossRef}]


\bibitem{Thooft:1972}
Hooft, G.; Veltman, M.
Regularization and Renormalization of Gauge Fields.
\emph{Nucl.\ Phys.\ B} {\bf 1972}, \emph{44}, 189--213. [\href{http://dx.doi.org/10.1016/0550-3213(72)90279-9}{CrossRef}]


\bibitem{Ashmore:1972}
Ashmore, J.F.
A Method of Gauge Invariant Regularization. \emph{Lett.\ Nuovo Cim.\ } {\bf 1972}, \emph{4}, 289--290. [\href{http://dx.doi.org/10.1007/BF02824407}{CrossRef}]


\bibitem{Cicuta:1972}
Cicuta, G.M.; Montaldi, E.
Analytic renormalization via continuous space dimension. \emph{Lett.\ Nuovo Cim.\ } {\bf 1972}, \emph{4}, 329--332. [\href{http://dx.doi.org/10.1007/BF02756527}{CrossRef}]


\bibitem{Wilson:1973}
Wilson, K.G.
Quantum field theory models in less than four-dimensions. \emph{Phys.\ Rev.\ D} {\bf 1973}, \emph{7}, 2911. [\href{http://dx.doi.org/10.1103/PhysRevD.7.2911}{CrossRef}]


\bibitem{Kislinger:1976}
Kislinger, M.B.; Morley, P.D.
Collective Phenomena in Gauge Theories. 2. Renormalization in Finite Temperature Field Theory. \emph{Phys.\ Rev.\ D} {\bf 1976}, \emph{13}, 2771. [\href{http://dx.doi.org/10.1103/PhysRevD.13.2771}{CrossRef}]
%



\bibitem{Donoghue:1983}
Donoghue, J.F.; Holstein, B.R.
Renormalization and Radiative Corrections at Finite Temperature.
\emph{\mbox{Phys.\ Rev.\ D}} {\bf 1983}, \emph{28}, 340, Erratum in {\bf 1984}, \emph{29}, 3004.  [\href{http://dx.doi.org/10.1103/PhysRevD.28.340}{CrossRef}]

\bibitem{Johansson:1986}
Johansson, A.E.I.; Peressutti, G.; Skagerstam, B.S.
Quantum Field Theory at Finite Temperature: Renormalization and Radiative Corrections.
\emph{Nucl.\ Phys.\ B} {\bf 1986}, \emph{278}, 324--342. [\href{http://dx.doi.org/10.1016/0550-3213(86)90216-6}{CrossRef}]


\bibitem{Kobes:1989}
Keil, W.; Kobes, R.
Mass and Wave Function Renormalization at Finite Temperature.
\emph{Phys. A} {\bf 1989}, \emph{158}, 47--57. [\href{http://dx.doi.org/10.1016/0378-4371(89)90506-2}{CrossRef}]


\bibitem{Keil:1989}
Keil, W.
Radiative Corrections and Renormalization at Finite Temperature: A Real Time Approach.
\emph{Phys.~Rev.~D} {\bf 1989}, \emph{40}, 1176. [\href{http://dx.doi.org/10.1103/PhysRevD.40.1176}{CrossRef}] [\href{http://www.ncbi.nlm.nih.gov/pubmed/10011926}{PubMed}]



\bibitem{LeBellac:1990}
Bellac, M.L.; Poizat, D.
Renormalization of External Lines in Relativistic Field Theories at Finite Temperature.
\emph{Z.\ Phys.\ C} {\bf 1990}, \emph{47}, 125--131. [\href{http://dx.doi.org/10.1007/BF01551922}{CrossRef}]


\bibitem{Elmfors:1992}
Elmfors, P.
Finite Temperature Renormalization of the $(\phi^3)_6$- and $(\phi^4)_4$-Models at Zero Momentum.
\mbox{\emph{Z.\ Phys.\ C}} {\bf 1992}, \emph{56}, 601. [\href{http://dx.doi.org/10.1007/BF01474733}{CrossRef}]


\bibitem{Eijck:1996}
Van Eijck, M.A.; Weert, C.G.V.
Finite-temperature renormalization of the phi**4(4) model. \emph{Int.\ J.\ Mod.\ Phys.\ B} {\bf 1996}, \emph{10}, 1485--1497. [\href{http://dx.doi.org/10.1142/S0217979296000593}{CrossRef}]


\bibitem{Chapman:1997}
Chapman, I.A. Finite temperature wave function renormalization: A Comparative analysis. \emph{Phys.\ Rev.\ D} {\bf 1997}, \emph{55}, 6287. [\href{http://dx.doi.org/10.1103/PhysRevD.55.6287}{CrossRef}]


\bibitem{Nakkagawa:1997}
Nakkagawa, H.; Yokota, H.
Effective potential at finite temperature: RG improvement versus high temperature expansion.
\emph{Prog.\ Theor.\ Phys.\ Suppl.\ } {\bf 1997}, \emph{129}, 209. [\href{http://dx.doi.org/10.1143/PTPS.129.209}{CrossRef}]



\bibitem{Baacke:1998}
Baacke, J.; Heitmann, K.; Patzold, C.
Renormalization of nonequilibrium dynamics at large N and finite temperature.  \emph{Phys.\ Rev.\ D} {\bf 1998}, \emph{57}, 6406. [\href{http://dx.doi.org/10.1103/PhysRevD.57.6406}{CrossRef}]


\bibitem{Esposito:1998}
Esposito, S.; Mangano, G.; Miele, G.; Pisanti, O.
Wave function renormalization at finite temperature.
\mbox{\emph{Phys.\ Rev.\ D}} {\bf 1998}, \emph{58}, 105023. [\href{http://dx.doi.org/10.1103/PhysRevD.58.105023}{CrossRef}]


\bibitem{Knoll:2002}
Van Hees, H.; Knoll, J.
Renormalization in selfconsistent approximation schemes at finite temperature. 3.~Global symmetries.
\emph{Phys.\ Rev.\ D} {\bf 2002}, \emph{66}, 025028. [\href{http://dx.doi.org/10.1103/PhysRevD.66.025028}{CrossRef}]


\bibitem{Jakovac:2005}
Jakovac, A.; Szep, Z.
Renormalization and resummation in finite temperature field theories.
\emph{Phys.\ Rev.\ D} {\bf 2005}, \emph{71}, 105001. [\href{http://dx.doi.org/10.1103/PhysRevD.71.105001}{CrossRef}]



\bibitem{Arrizabalaga:2007}
Arrizabalaga, A.; Reinosa, U.
Renormalized finite temperature phi**4 theory from the 2PI effective action.
\emph{Nucl.\ Phys.\ A} {\bf 2007}, \emph{785}, 234. [\href{http://dx.doi.org/10.1016/j.nuclphysa.2006.11.143}{CrossRef}]


\bibitem{Blaizot:2007}
Blaizot, J.P.; Ipp, A.; Mendez-Galain, R.; Wschebor, N.
Perturbation theory and non-perturbative renormalization flow in scalar field theory at finite temperature.
\emph{Nucl.\ Phys.\ A} {\bf 2007}, \emph{784}, 376--406. [\href{http://dx.doi.org/10.1016/j.nuclphysa.2006.11.139}{CrossRef}]


\bibitem{Blaizot:2015}
Blaizot, J.P.; Wschebor, N.
Massive renormalization scheme and perturbation theory at finite temperature.
\emph{Phys.\ Lett.\ B} {\bf 2015}, \emph{741}, 310--315. [\href{http://dx.doi.org/10.1016/j.physletb.2014.12.040}{CrossRef}]


\bibitem{RyderBook}
Ryder, L.H. \emph{Quantum Field Theory}; Cambridge University Press: Cambridge, UK, {1985}.


\bibitem{Garbrecht:2002pd}
Garbrecht, B.; Prokopec, T.; Schmidt, M.G.
Particle number in kinetic theory. \emph{Eur.\ Phys.\ J.\ C} {\bf 2004}, \emph{38}, 135--143. [\href{http://dx.doi.org/10.1140/epjc/s2004-02007-0}{CrossRef}]

\bibitem{Dadic:1998yd}
Dadić, I. Two mechanisms for elimination of pinch singularities in/out of equilibrium thermal field theories.
\emph{Phys.\ Rev.\ D} {\bf 1999}, \emph{59}, 125012. [\href{http://dx.doi.org/10.1103/PhysRevD.59.125012}{CrossRef}]


\end{thebibliography}
\end{document}